\documentclass[%
reprint,
amsmath,
amssymb,
aps,
prl,
floatfix,
]{revtex4-2}

\usepackage{graphicx}
\usepackage{dcolumn}
\usepackage{bm}
\usepackage[hypertexnames=false]{hyperref}
\hypersetup{
	colorlinks = true,
	allcolors = blue,
}
\usepackage{multirow}

\usepackage{siunitx}

\usepackage{amsmath, amssymb}
\usepackage{bbold}

\newcommand{\Mi}{\mathrm{i}}
\newcommand{\Me}{\mathrm{e}} 
\newcommand{\Mbf}[1]{\boldsymbol{#1}} 
\newcommand{\Mdiff}{\mathrm{d}} 

\renewcommand{\vec}[1]{\Mbf{#1}}

\newcommand{\tp}{{\mathrm{T}}}

\newcommand{\x}{\vec{x}}
\newcommand{\xxi}{\vec{\xi}}

\begin{document}

\title{Efficient matter-wave lensing of ultracold atomic mixtures}

\author{Matthias Meister}
\email[Corresponding author. Email: ]{matthias.meister@dlr.de}

\author{Albert Roura}

\affiliation{Institute of Quantum Technologies, German Aerospace Center (DLR), Wilhelm-Runge-Straße 10, 89081 Ulm, Germany}

\date{\today}

\begin{abstract}

Mixtures of ultracold quantum gases are at the heart of high-precision quantum tests of the weak equivalence principle, where extremely low expansion rates have to be reached with matter-wave lensing techniques. We propose to simplify this challenging atom-source preparation by employing magic laser wavelengths for the optical lensing potentials which guarantee that all atomic species follow identical trajectories and experience common expansion dynamics. In this way, the relative shape of the mixture is conserved during the entire evolution while cutting in half the number of required lensing pulses compared to standard approaches. 
\end{abstract}

\maketitle

%%%%%%%%%%%%%%%%%%%%%%%%%%%%%%%%%%%%%%%%%%%%%%%%%%%
%%%%%%%%%%%%%%%%%%%%%%%%%%%%%%%%%%%%%%%%%%%%%%%%%%%
% 
%					INTRODUCTION
% 
%%%%%%%%%%%%%%%%%%%%%%%%%%%%%%%%%%%%%%%%%%%%%%%%%%%
%%%%%%%%%%%%%%%%%%%%%%%%%%%%%%%%%%%%%%%%%%%%%%%%%%%

%\section*{}
\subsection{Introduction}

Mixtures of ultracold quantum gases are unique systems to study a plethora of phenomena including molecule formation~\cite{Koehler_2006,Carr_2009} and many-body physics~\cite{Bloch_2008} with different isotopes and elements, the realization of shell-shaped quantum gases based on interspecies repulsion~\cite{Wolf_2022}, as well as testing the universality of free fall (UFF)~\cite{Schlippert_2014,Rosi_2017,Asenbaum2020}, where the gravitational acceleration of two different atomic species is compared by means of atom interferometric techniques~\cite{Cronin2009}. 
Nowadays different types of atomic mixtures are studied in this context ranging from laser-cooled thermal clouds, Bose-Einstein condensates (BECs)~\cite{Dalfovo1999}, and degenerate Fermi gases~\cite{Giorgini_2008} to combinations of ultracold bosons and fermions. Despite their obvious differences, all these systems share similar experimental challenges due to the nature of a quantum gas mixture.

\begin{figure*}
\includegraphics[scale=0.8]{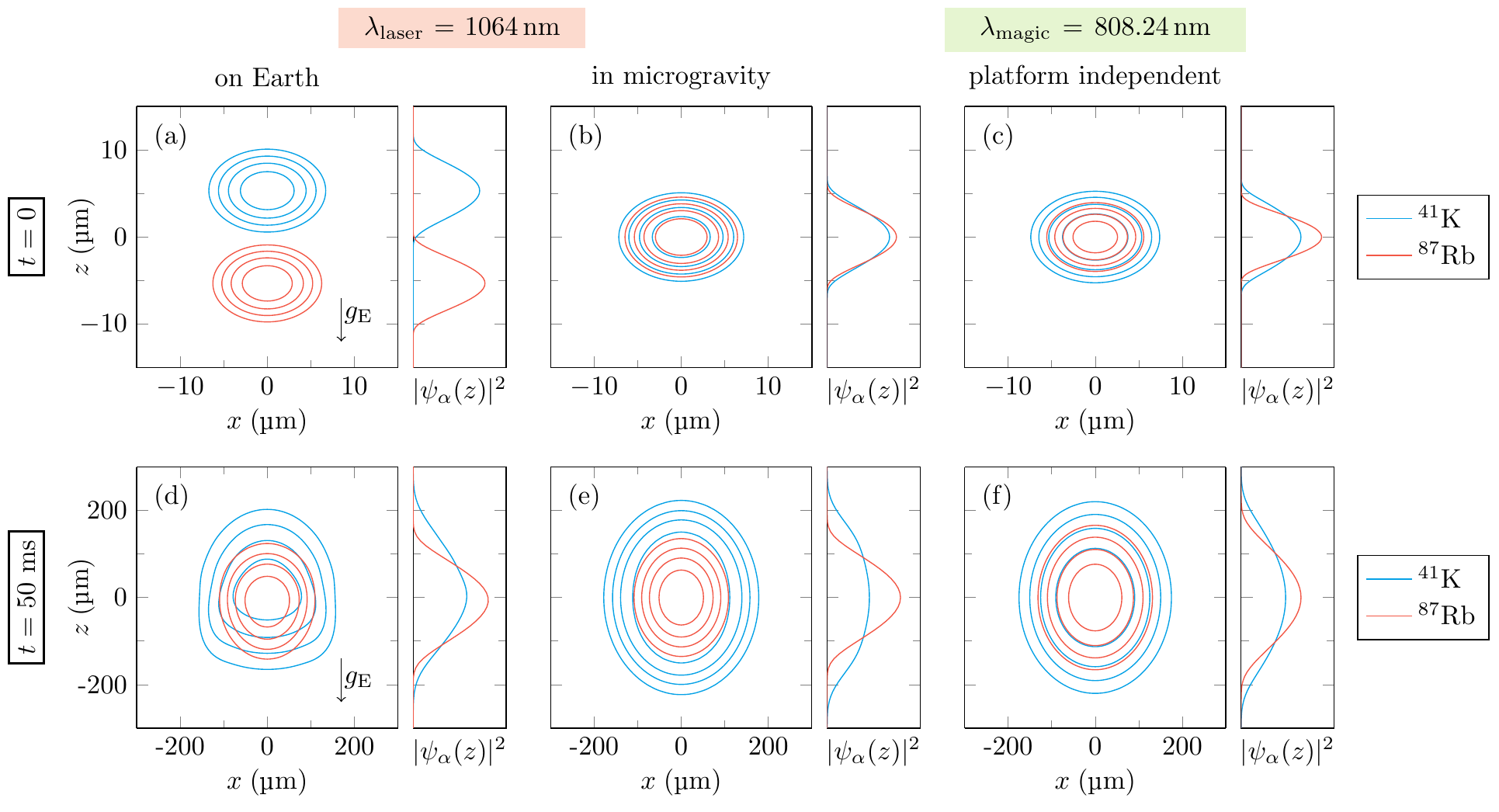}
\caption{Integrated 2D density distributions in the comoving $x$-$z$-plane of an interacting $^{41}$K-$^{87}$Rb BEC mixture displaying the ground state (a--c) and the state after $50$ \si{\milli\second} of free expansion (d--f) for three combinations of laser wavelengths and gravitational environments. The side panels depict the integrated 1D density along $z$.
On Earth and in the typical case of a non-magic wavelength $\lambda_\mathrm{laser} = 1064$ \si{\nano\meter} the gravitational acceleration $g_\mathrm{E} = 9.81\,\si{\meter\second^{-2}}$ leads to a relative displacement between the two condensates, potentially preventing any overlap for the ground state (a) and resulting in asymmetric and deformed densities after some time of flight (d). 
In microgravity the ground state is symmetric (b) but its relative shape changes after release from the trap (e) as a consequence of the different trap frequencies for both species ($\omega_\mathrm{Rb} = 0.733 \,\omega_\mathrm{K}$). 
In contrast, for the magic wavelength $\lambda_\mathrm{magic} = 808.24$ \si{\nano\meter}, which yields $\omega_\mathrm{K} = \omega_\mathrm{Rb}$, the ground state is symmetric (c) and in addition fully conserves its relative shape during free expansion (f).
The parameters for these exact numerical simulations are the trap frequency $\omega_\mathrm{K} = 2\pi\cdot 100$ \si{\Hz}, the number of particles $N_\mathrm{K} = N_\mathrm{Rb} = 10^5$, and the $s$-wave scattering lengths $a_{\mathrm{KK}} = 60 a_0$, $a_{\mathrm{RbRb}} = 100 a_0$, and $a_{\mathrm{KRb}} = 20 a_0$. 
\label{fig:ground_state_free_expansion}}
\end{figure*}

For instance on Earth, the gravitational sag typically leads to asymmetric states~\cite{Myatt_1997,Modugno_2002,Riboli_2002} due to different central trajectories of the individual species as shown in the first column of Fig.~\ref{fig:ground_state_free_expansion} for the particular case of a $^{41}$K-$^{87}$Rb BEC mixture. This differential sag limits the efficiency of sympathetic cooling as well as the precision of UFF tests~\cite{Aguilera_2014,Battelier_2021} and prevents the formation of shell-structures for mixtures on ground~\cite{Wolf_2022}.

A possible way of circumventing unwanted effects due to gravitational sag is the operation on microgravity platforms such as drop towers~\cite{Muentinga2013,Condon_2019}, zero-g planes~\cite{Barrett_2016} or in space~\cite{Becker_2018,Aveline_2020,Frye_2021}. However, even then the free expansion rates of the different species do not match for the trap configurations being considered in standard setups; see middle column of Fig.~\ref{fig:ground_state_free_expansion}. As a consequence, the relative density distributions of the mixture are deformed by the expansion dynamics, spoiling symmetric shell structures and leading to a decrease in precision for interferometry measurements due to relative wave-front distortions. 

There are methods to engineer the expansion rates of the mixture by switching on a trapping potential that acts as a matter-wave lens for a short time after the clouds have expanded. This delta-kick collimation technique~\cite{Chu1986,Ammann1997,Muentinga2013,Kovachy2015,Corgier_2020,Deppner_2021,Gaaloul_2022} reduces the kinetic energy of the system and enables very low expansion rates, which are mandatory for high-precision interferometry in microgravity. However, in order to control the expansion rates of a mixture in all spatial dimensions one lens per independent direction is required for each species, rendering a full 3D implementation rather impractical and error prone.

In this article we propose to overcome these challenges by taking advantage of optical dipole traps employing special \textit{magic} wavelengths so that the ratio of the optical potentials for two different species $j$ and $k$ is given by the ratio of their masses: $V_j / V_k = m_j / m_k$~\footnote{Note that for different internal states of the same atomic isotope the masses $m_j$ and $m_k$ are identical and one recovers the usual notion of ``magic wavelength'' employed in optical atomic clocks~\cite{Ludlow2015}.}. 
In this case, the classical equations of motion for the center-of-mass dynamics are identical for all species, which results in a vanishing differential gravitational sag~\cite{Ospelkaus2006} and perfect co-location of the species involved even in an Earth-based laboratory; see Fig.~\ref{fig:ground_state_free_expansion}~c.

Most importantly, as we will show here, optical dipole traps with such a magic laser wavelength guarantee that the relative shape of the mixture is conserved during the entire dynamics even for time-dependent potentials, see last column of Fig.~\ref{fig:ground_state_free_expansion}. This property has far-reaching consequences and for instance allows to efficiently collimate mixtures of ultracold atoms to reach very low expansion rates
with fewer lensing pulses than standard approaches. 

Hence, employing the magic wavelengths proposed here will drastically improve the control over the dynamics of mixture experiments, allowing new applications in ground-based setups, but also boosting long-time dual-species interferometry in space by simplifying the source preparation and at the same time minimizing spurious phase shifts due to wave-front distortions that would otherwise undermine the accuracy of the measurements.

The proposed method applies to several types of mixtures, namely thermal gases, BECs, and non-interacting ultracold Fermi gases, 
but not for Bose-Fermi mixtures. As a specific example we focus here on the particularly interesting case of a multi-species BEC.

%%%%%%%%%%%%%%%%%%%%%%%%%%%%%%%%%%%%%%%%%%%%%%%%%%%
%%%%%%%%%%%%%%%%%%%%%%%%%%%%%%%%%%%%%%%%%%%%%%%%%%%
% 
%				DYNAMICS OF MIXTURES
% 
%%%%%%%%%%%%%%%%%%%%%%%%%%%%%%%%%%%%%%%%%%%%%%%%%%%
%%%%%%%%%%%%%%%%%%%%%%%%%%%%%%%%%%%%%%%%%%%%%%%%%%%

\subsection{Common dynamics for quantum gas mixtures}

As long as the trapping potentials of the different atomic species fulfill the condition $V_j / V_k = m_j / m_k$, which can be achieved by employing a magic laser wavelength and also holds for gravitational and inertial forces, 
the classical equations of motion for the center-of-mass are identical for all species and we can describe the mixture with a comoving coordinate system that follows the classical trajectory; see the \hyperlink{sec:appendix_hyperlink}{Appendix}.
In this comoving frame and for locally harmonic potentials $V_j$, the expansion dynamics is therefore governed by potentials of the form
\begin{equation}\label{eq:trap_potential}
 \mathcal{V}_{j}(\x,t) / m_j = \x^\tp \Omega^2(t) \,\x / 2
\end{equation}
where $m_j$ is the atom mass of species $j$ and $\Omega^2(t)$ is a $3 \times 3$ matrix containing the trap frequencies. 

Since the local trap frequencies of potentials given by Eq.~\eqref{eq:trap_potential} are equal for all species, we can describe the time evolution of the mixture with a common scaling that conserves the relative shape of the density distributions analogously to the single-species case~\cite{Castin1996,Kagan1996,Meister_2017}.
Indeed, as shown in more detail in the \hyperlink{sec:appendix_hyperlink}{Appendix}, starting with the well-known nonlinear Gross-Pitaevskii equation (GPE), the transformation
\begin{equation}\label{eq:transformation_wavefunction}
  \psi_j(\x,t) = \Me^{\Mi\,\Phi_j(\xxi,t)} \, \psi_{\Lambda,j}\left(\xxi, t\right) / \sqrt{\det\Lambda}
\end{equation}
of the wave function to the adapted coordinates $\xxi = \Lambda^{-1}(t) \,\x$ with the quadratic phase $\Phi_j(\xxi,t)$ leads to a transformed GPE with nearly vanishing time evolution for the wave function $\psi_{\Lambda,j}(\xxi,t)$.

The reason for this nearly frozen dynamics in the adapted coordinates is the fact that the position-dependent quadratic phase $\Phi_j(\xxi,t)$ already contains the entire dynamics as long as the time-dependent Thomas-Fermi approximation is valid and locally harmonic potentials are considered. 
In this case, the relative shape of the wave function does not change and the time evolution in the comoving frame is fully determined by the transformation in Eq.~\eqref{eq:transformation_wavefunction} and the scaling matrix $\Lambda = \Lambda(t)$, which is a time-dependent $3\times 3$ matrix that fulfills the differential equation
\begin{equation}\label{eq:diff_equation_Lambda}
 \frac{\Mdiff^2\Lambda}{\Mdiff t^2} + \Omega^2(t) \, \Lambda = \frac{\left(\Lambda^{-1}\right)^\tp  \Omega^2(0)}{\det\Lambda}
\end{equation}
with the initial conditions $\Lambda(0) = \mathbb{1}$ and $\dot\Lambda(0) = 0$.

\begin{figure*}
\includegraphics[scale=0.8]{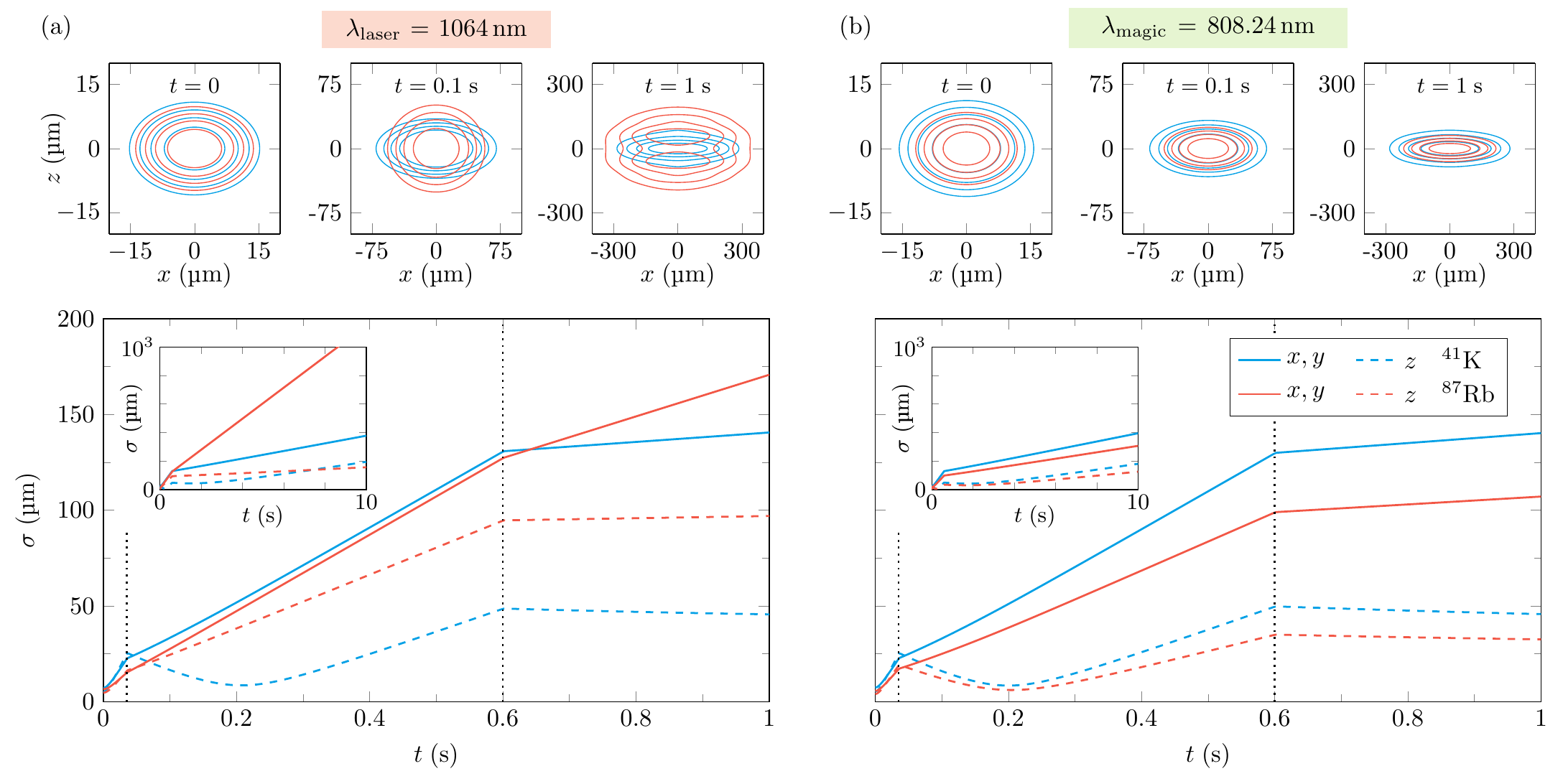}
\caption{Time evolution of the density distributions (top) and standard deviations $\sigma(t)$ (bottom) of an interacting $^{41}$K-$^{87}$Rb BEC mixture for a sequence of two atomic-lensing pulses. 
During free expansion the trapping potential is switched on again for a first (second) lens pulse $35\;(600)$ \si{\milli\second} after release from the trap for a duration of $1.85\;(2.40)$ \si{\milli\second} with frequencies scaled by a factor of $1\;(1/4)$. 
For the standard laser wavelength $\lambda_\mathrm{laser} = 1064$ \si{\nano\meter} (a) the lenses act differently on both species ($\omega_\mathrm{Rb} = 0.733 \,\omega_\mathrm{K}$) resulting in an almost uncollimated expansion of the rubidium cloud (red) along the $x$- and $y$-axis deforming the whole mixture due to interspecies interactions. 
On the other hand, for the magic wavelength $\lambda_\mathrm{magic} = 808.24$ \si{\nano\meter} (b) simultaneous collimation for both species in all directions is naturally accomplished ($\omega_\mathrm{Rb} = \,\omega_\mathrm{K}$), the relative shape of the mixture is conserved during the entire evolution and cloud sizes below $400$ \si{\micro\meter} for up to $10$ \si{\second} are attained (see inset), which corresponds to 3D expansion energies of $3.4$ and $4.3$ \si{\pico\kelvin}.
The cylindrical symmetric setup for this numerical simulation is defined by the radial trap frequency $\omega_\mathrm{K} = 2\pi\cdot 15$ \si{\Hz}, the number of particles $N_\mathrm{K} = N_\mathrm{Rb} = 10^5$, and the $s$-wave scattering lengths $a_{\mathrm{KK}} = 60 a_0$, $a_{\mathrm{RbRb}} = 100 a_0$, and $a_{\mathrm{KRb}} = 20 a_0$.
\label{fig:size_evolution_DKC}}
\end{figure*} 

We emphasize that a constant, non-vanishing interspecies interaction does not change the relative shape of the mixture during the dynamics for potentials satisfying Eq.~\eqref{eq:trap_potential} because the effect of the interaction has already been taken into account for the initial ground state of the mixture which is simply rescaled by the expansion dynamics as shown in the right column of Fig.~\ref{fig:ground_state_free_expansion}.

%%%%%%%%%%%%%%%%%%%%%%%%%%%%%%%%%%%%%%%%%%%%%%%%%%%
%%%%%%%%%%%%%%%%%%%%%%%%%%%%%%%%%%%%%%%%%%%%%%%%%%%
% 
%					ATOMIC LENSING
% 
%%%%%%%%%%%%%%%%%%%%%%%%%%%%%%%%%%%%%%%%%%%%%%%%%%%
%%%%%%%%%%%%%%%%%%%%%%%%%%%%%%%%%%%%%%%%%%%%%%%%%%%

\subsection{Efficient atomic lensing for mixtures}

Since our method works for time-dependent potentials which are locally harmonic, it can also be exploited for matter-wave lensing of the ultracold atomic mixture with delta-kick collimation techniques~\cite{Chu1986,Ammann1997,Muentinga2013,Kovachy2015,Corgier_2020,Deppner_2021,Gaaloul_2022}.
In order to achieve 3D lensing in general one lens pulse is required for each independent direction, which for a single quantum gas in a cylindrically symmetric trap implies two pulses and four for a dual-species mixture.

By employing magic wavelengths for crossed optical dipole traps, which lead to local potentials given by Eq.~\eqref{eq:trap_potential} with a diagonal matrix $\Omega^2(t) = \omega^2(t)\,\mathrm{diag}(1,1,2)$, 
this requirement can be reduced to only two pulses for a dual-species mixture because the relative shape of the mixture is conserved by the time evolution and both species respond equally to the lensing potential. 

As a relevant example, in Fig.~\ref{fig:size_evolution_DKC} we compare the performance of a double-lens sequence acting on a $^{41}$K-$^{87}$Rb BEC mixture for the cases of standard and magic laser wavelengths. During the free expansion of the mixture a first lensing pulse is applied 35 \si{\milli\second} after release from the trap, followed by a second lensing pulse at $t = 600$ \si{\milli\second} such that very low expansion rates are achieved for $^{41}$K in all spatial directions. In addition, the interspecies scattering length is tuned to the constant value $a_{\mathrm{KRb}} = 20 a_0$, where $a_0$ is the Bohr radius, by use of a Feshbach resonance~\cite{Thalhammer_2008} resulting in a large spatial overlap of the ground state densities of the two species. All chosen parameters are realistic for operation in microgravity and are aligned with the requirements of future space missions for precision measurements~\cite{Battelier_2021}.

As shown in Fig.~\ref{fig:size_evolution_DKC}~b, for the magic laser wavelength an almost perfect collimation can be achieved and the size of the clouds along all three spatial directions remain below 400 \si{\micro\meter} for an expansion time of 10 \si{\second}, which corresponds to theoretically achievable 3D expansion energies of 3.4 and 4.3 \si{\pico\kelvin} for $^{41}$K and $^{87}$Rb, respectively.
Moreover, when employing a magic laser wavelength the relative shape is fully conserved during the entire evolution and irrespective of the constant interspecies scattering length.

In contrast, in the case of a standard laser wavelength, displayed in Fig.~\ref{fig:size_evolution_DKC}~a, the initially inner species ($^{87}$Rb) expands much faster after the lensing pulses (see inset for long times), which leads to strong deformations of the density distribution of both species that are enhanced by the interspecies interaction. Such behavior can lead to spurious interferometric phases that can severely compromise the measurement accuracy.
Furthermore, since good collimation in all directions is only achieved for $^{41}$K,  the rather large expansion rate along $x$ and $y$ for $^{87}$Rb would ultimately limit the time that the atom cloud could be potentially observed. 
The relatively slow expansion of $^{87}$Rb in the $z$-direction is due to a fortunate coincidence because the frequency in this direction is very close to the one of $^{41}$K in $x$ and $y$, which is due to the mass ratio and polarizability of the two species.  

In summary, the common translation and expansion dynamics for mixtures of ultracold atoms afforded by magic laser wavelengths is a very natural way of improving the efficiency of atomic lensing protocols which are mandatory for precision UFF tests and other applications such as studying the dynamics of quantum bubbles based on mixtures without compromising the shell structure during the free expansion.

%%%%%%%%%%%%%%%%%%%%%%%%%%%%%%%%%%%%%%%%%%%%%%%%%%%
%%%%%%%%%%%%%%%%%%%%%%%%%%%%%%%%%%%%%%%%%%%%%%%%%%%
% 
%			OPTIMAL LASER WAVELENGTHS
% 
%%%%%%%%%%%%%%%%%%%%%%%%%%%%%%%%%%%%%%%%%%%%%%%%%%%
%%%%%%%%%%%%%%%%%%%%%%%%%%%%%%%%%%%%%%%%%%%%%%%%%%%

\subsection{Magic laser wavelengths}

For the typical example of a crossed optical dipole trap~\cite{Grimm2000} discussed above, the square of the radial trap frequency is given by 
\begin{equation}\label{eq:trap_frequency_CODT}
 \omega_j^2 = 4 \,P_\mathrm{L} \,\mathrm{Re}\left[\alpha_{\mathrm{L},j}\right] / \left(\pi \,c \,\epsilon_0 \,w_\mathrm{L}^4 \,m_j\right)
\end{equation}
where $P_\mathrm{L}$ is the laser power, $w_\mathrm{L}$ the beam waist and $\alpha_{\mathrm{L},j}$ the polarizability, which includes the contributions from all relevant atomic transitions.

\begin{figure}
\includegraphics[scale=0.8]{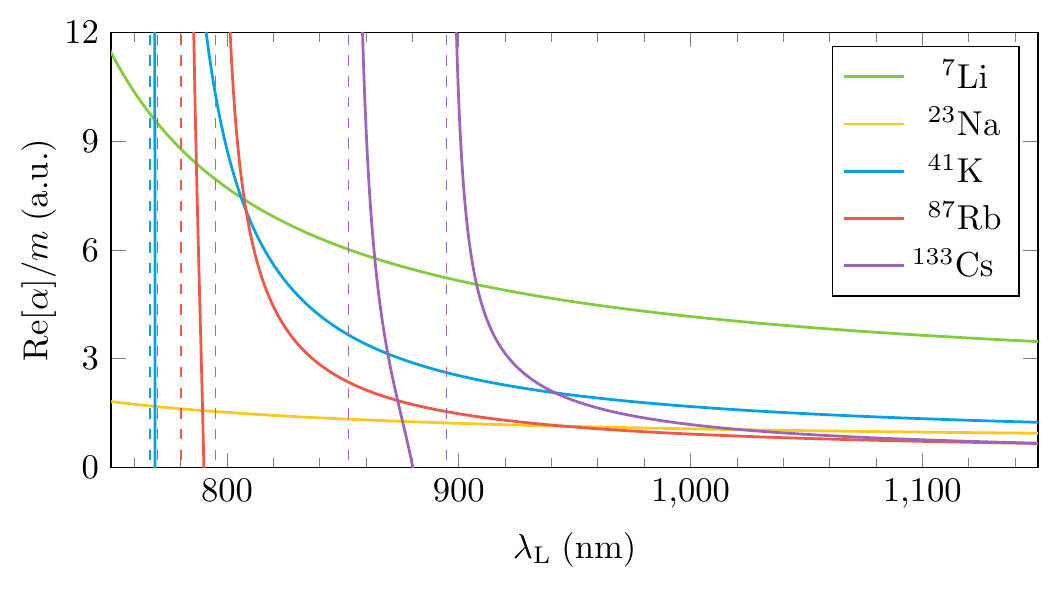}
\caption{Polarizability $\alpha$ divided by the atom mass $m$ as a function of the laser wavelength $\lambda_\mathrm{L}$ for all stable alkaline elements. Line crossings indicate magic laser wavelengths yielding a common dynamics for the two species involved. The resonant D-transition lines are marked by vertical dashed lines. 
\label{fig:trap_frequencies_over_laser_wavelength}}
\end{figure}

From Eq.~\eqref{eq:trap_frequency_CODT} we see that the ratio between the polarizability and the mass contains the entire species dependence of the trap frequency. In Fig.~\ref{fig:trap_frequencies_over_laser_wavelength} this ratio is plotted as a function of the laser wavelength $\lambda_\mathrm{L}$ for all alkali metals. Similar calculations for the polarizability have been done before in Refs.~\cite{Safronova_2006,MeisterPhD2019}.
As clearly seen in Fig.~\ref{fig:trap_frequencies_over_laser_wavelength}, there are several crossing points for different pairs of elements, which correspond to magic wavelengths leading to the common dynamics for both species discussed in the present paper.

In particular binary combinations of $^{23}$Na, $^{87}$Rb, and $^{133}$Cs feature crossing points far away from their transition frequencies and are therefore excellent candidates for implementing our proposed scheme. 
Of special interest is also the crossing of $^{41}$K and $^{87}$Rb at $\lambda_\mathrm{L} = 808.24\,\si{\nano\meter}$ because this mixture is considered a prime candidate for future space-based tests of UFF~\cite{Battelier_2021}. 
In fact, a similar wavelength has been used in Ref.~\cite{Ospelkaus2006} to avoid the differential gravitational sag of a $^{40}$K-$^{87}$Rb mixture. However, for such a Bose-Fermi mixture there is no common expansion dynamics.

%%%%%%%%%%%%%%%%%%%%%%%%%%%%%%%%%%%%%%%%%%%%%%%%%%%
%%%%%%%%%%%%%%%%%%%%%%%%%%%%%%%%%%%%%%%%%%%%%%%%%%%
% 
%					DISCUSSION
% 
%%%%%%%%%%%%%%%%%%%%%%%%%%%%%%%%%%%%%%%%%%%%%%%%%%%
%%%%%%%%%%%%%%%%%%%%%%%%%%%%%%%%%%%%%%%%%%%%%%%%%%%

\subsection{Discussion}

In order to assess the practical feasibility of magic laser wavelengths, the required laser power and atom losses due to spontaneous emission need to be considered.
For the particular lensing sequence of a $^{41}$K-$^{87}$Rb BEC mixture with the magic laser wavelength $\lambda_\mathrm{L} = 808.24\,\si{\nano\meter}$ discussed in this paper (see Fig.~\ref{fig:size_evolution_DKC}~b) a beam waist of $100\;(600)\,\si{\micro\meter}$ and a laser power of $2.61\;(212)\,\si{\milli\watt}$ are required for the first (second) lens pulse, respectively. 
These parameters constitute a reduction by a factor of 5 in laser power compared with a setup based on the typical laser wavelength $\lambda_\mathrm{L} = 1064\,\si{\nano\meter}$. 
Hence, for currently planned high-precision UFF tests in space our approach can substantially relax the requirements on the needed laser power in addition to making the lensing process much more efficient as discussed in the previous sections. The required laser light can be generated with standard diode lasers at 808 nm, which are already available for ground operations and can be further qualified for space missions.
Moreover, atom losses due to off-resonant photon scattering are negligible during the rather short lensing pulses, where peak scattering rates of $7.77 (17.5) \cdot 10^{-3}$ 1/s for $^{41}$K and $35.0 (78.7) \cdot 10^{-3}$ 1/s for $^{87}$Rb for the first (second) lens occur for $\lambda_\mathrm{L} = 808.24\,\si{\nano\meter}$. For these scattering rates, performing the final evaporation in the optical trap seems feasible as well.

When comparing our approach to another recent proposal for atomic lensing of BEC mixtures~\cite{Corgier_2020}, we identify three main improvements that result from using a magic wavelength: (i) the common dynamics enabled by the magic laser wavelength allows a reduction of the required number of lens pulses up to a factor of 2, (ii) less laser power is needed due to working with a smaller detuning, and (iii) the interspecies interaction does not need to be switched to zero during the lens sequence in order to avoid deformations of the clouds, and the scattering length can instead remain constant with a non-vanishing value during the whole process. 
For the $^{41}$K-$^{87}$Rb mixture discussed in detail above, adjusting the interspecies interaction through a Feshbach resonance~\cite{Thalhammer_2008} is nevertheless required in order to obtain a miscible initial state. The large magnetic field that is applied to address the Feshbach resonance needs to be switched off after the entire lens sequence since it would disturb the interferometric measurement. By optimizing the timings of the sequence, the impact of this late change of the interspecies interaction on the expansion dynamics can be made negligible.
Alternatively, by considering mixtures with a low natural interspecies scattering length and a miscible ground state, such as $^{84}$Sr-$^{86}$Sr~\cite{Stellmer2013}, there would be no need to use a Feshbach field at all, which further relaxes the complexity of the atom-source preparation.

Besides these crucial advantages for matter-wave lensing of ultracold atomic mixtures, there are also other applications that will benefit from magic optical wavelengths such as the generation in ground-based laboratories of shell-shaped BEC mixtures, which can be exploited to investigate curved 2D-geometries in 3D space, or the study of molecule formation on ground due to an improved overlap of the two different species involved. Lastly, our study of magic wavelengths (Fig.~\ref{fig:trap_frequencies_over_laser_wavelength}) has also revealed the possibility of a three-species mixture of $^7$Li, $^{41}$K, and $^{87}$Rb with large spatial overlap in Earth-bound experiments that could open up new avenues in many-body physics.

\bibliography{bibliography}

%%%%%%%%%%%%%%%%%%%%%%%%%%%%%%%%%%%%%%%%%%%%%%%%%%%
%%%%%%%%%%%%%%%%%%%%%%%%%%%%%%%%%%%%%%%%%%%%%%%%%%%
% 
%					Appendix
% 
%%%%%%%%%%%%%%%%%%%%%%%%%%%%%%%%%%%%%%%%%%%%%%%%%%%
%%%%%%%%%%%%%%%%%%%%%%%%%%%%%%%%%%%%%%%%%%%%%%%%%%%

\clearpage

\onecolumngrid

\section{Appendix: \\ Evolution of BEC mixtures in optical potentials for magic laser wavelengths}\label{sec:appendix_mixture_dynamics}
\hypertarget{sec:appendix_hyperlink}{}

In this appendix we show that in the case of magic laser wavelengths and for locally harmonic potentials the central trajectories of all atomic species are identical for quantum gas mixtures and that the evolution of the centered wave packets can be described by a single scaling in the case of BEC mixtures. 

\subsection{Classical equations of motion}

From the classical equation of motion $m\, \ddot{\vec{x}} = - \boldsymbol{\nabla} V$ it is clear that when the ratio $V_j / m_j$ is the same for different atomic species (labeled here with the subindex $j$), both the equations and the corresponding solutions will be the same for all species. Furthermore, given any external potentials with this property, 
such as the optical potentials for magic laser wavelengths considered in this article, the same conclusions will also apply in the presence of a gravitational field, as can be immediately seen from the equation of motion in that case:
\begin{equation}
m_j \ddot{\vec{x}} = - \boldsymbol{\nabla} \big( V_j(\vec{x},t) + m_j U(\vec{x},t) \big)
\label{eq:classical_eom} ,
\end{equation}
where $U(\vec{x},t)$ is the gravitational potential.
In addition, the result can be straightforwardly generalized to include the effects of inertial forces as well (for example, due to accelerations and rotations of the experimental setup) and it leads in all cases to identical trajectories for the different atomic species.

In particular, this implies that the gravitational sag for a trap potential, which can be determined by taking $\ddot{\vec{x}} = 0$ on the left-hand side of Eq.~\eqref{eq:classical_eom}, will be the same for all species. Moreover, the  frequency matrix $(\Omega^2)_{ab} = (1/m_j)\, \partial^2 V_j / \partial x_a \partial x_b$ at the trap minimum will be common for all atomic species and, more generally, this will also hold for the local 
frequency matrix obtained by evaluating the second derivatives of the potential at any other position.
As a simple example, if we consider a harmonic external potential in a uniform gravitational field, corresponding to $U(\vec{x}) = U_0 - \vec{g} \cdot \vec{x}$, and with one of its principal axis aligned with the gravitational acceleration $\vec{g}$, the resulting gravitational sag is $\Delta z = - g / \omega_z^2$. Here $\omega_z^2$ is the eigenvalue of the matrix $\Omega^2$, which in this case is spatially independent, along the direction of the gravitational field.
For non-uniform gravitational fields gravity gradients lead to non-vanishing second derivatives of the gravitational potential $U$, whose contribution can be included in the frequency matrix $\Omega^2$. In practice, however, such contributions are much smaller than those from the external potential $V_j$ and can typically be neglected.

\subsection{Propagation of matter wave packets}

Interestingly, the above conclusions for the classical case can be naturally extended to the quantum dynamics of matter wave packets and atomic clouds. In order to show this point, it is particularly useful to consider a description of matter wave propagation in terms of \emph{central trajectories} and \emph{centered wave packets} \cite{Borde2001,Hogan2008,Roura2014,Meister_2017} which is applicable to a very broad range of situations, including a relativistic description of matter wave propagation in curved spacetime \cite{Roura2020}.
Further details can be found in the quoted references, but the key result is that the wave-packet evolution can be expressed as follows:
\begin{equation}
\psi_j (\vec{x},t) = \Me^{\Mi\,  S_j / \hbar} \, \Me^{\Mi\,  \vec{P}(t) \cdot \left( \vec{x} - \vec{X}(t) \right) / \hbar} \,
\psi^{\text(c)}_j \big( \vec{x} - \vec{X}(t), t \big)
\label{eq:wave-packet_evolution} ,
\end{equation}
where $\vec{X}(t)$ and $\vec{P}(t) = m_j \dot{\vec{X}}(t)$ correspond to the central trajectory and satisfy the classical equations of motion, $\psi^{\text(c)}_j$ is the centered wave function, and the phase $S_j$ is given by the classical action
\begin{equation}
S_j = \int^{t}_{t_0} dt' \left( \frac{1}{2} m_j \dot{\vec{X}}^2 - V_j (\vec{X}, t')
- m_j\,  U (\vec{X}, t') \right)
\label{eq:phase} .
\end{equation}
More importantly for our considerations in the present paper, the evolution of the centered wave packet is governed by the following Schr\"odinger equation:
\begin{equation}
\Mi \hbar \,\frac{\partial}{\partial t} \,\psi^{\text(c)}_j(\x',t) = \left[-\frac{\hbar^2}{2m_j} \nabla_{\x'}^2
+ \mathcal{V}_j(\x',t) \right] \psi^{\text(c)}_j(\x',t)
\label{eq:centered_wave-packet_evolution} ,
\end{equation}
where we have introduced the comoving coordinate $\vec{x}' = \vec{x} - \vec{X}$  and $\mathcal{V}_j (\vec{x}',t) \equiv V_j (\vec{X} + \vec{x}',t) - V_j (\vec{X},t) - \vec{x}' \cdot \boldsymbol{\nabla} V_j (\vec{X},t)$, which reduces to the purely quadratic part when $V_j$ is a harmonic potential.
For non-uniform fields the gravitational potential $U$ gives rise to a contribution analogous to $\mathcal{V}_j$, but it is typically much smaller and has been omitted here.
Note also that in the main text, e.g.\ in Eqs.~\eqref{eq:trap_potential}--\eqref{eq:transformation_wavefunction} and Figs.~\ref{fig:ground_state_free_expansion}--\ref{fig:size_evolution_DKC}, we have used $\vec{x}$ instead of $\vec{x}'$ for the comoving coordinates in order to ease the notation.

For BECs and BEC mixtures, one needs to add the mean-field interaction term and obtains then the Gross-Pitaevskii equation (GPE) for the centered wave functions describing the multispecies condensate in the mean-field approximation:
\begin{equation}
\Mi \hbar \,\frac{\partial}{\partial t} \,\psi^{\text(c)}_j(\x',t)   = \left[-\frac{\hbar^2}{2m_j} \nabla_{\x'}^2 + \mathcal{V}_j(\x',t) + \sum\limits_k g_{jk} \left|\psi^{\text(c)}_k(\x',t)\right|^2 \right] \psi^{\text(c)}_j(\x',t)
\label{eq:gpe} ,
\end{equation}
where the subindex $j$ labels the atomic species and we have taken into account that for the potentials being considered the central trajectory $\vec{X}(t)$ is the same for all species. If in addition the potential can be regarded as locally harmonic (i.e.\ well approximated by a harmonic potential over the size of the atomic cloud), in Eq.~\eqref{eq:gpe} one can simply take
\begin{equation}
\mathcal{V}_j (\vec{x}',t) = \frac{m_j}{2} \vec{x}'^\tp \Omega^2(t) \,\vec{x}' \label{eq:harmonic_potential} ,
\end{equation}
with a common frequency matrix $\Omega^2 (t)$ for all species and corresponding to the second derivatives of the potential evaluated at the central trajectory: $(\Omega^2)_{ab}(t) = (1/m_j) \left. \partial^2 V_j / \partial x_a \partial x_b \right |_{\vec{x} = \vec{X}(t)}$.

The initial ground state of the mixture is in this case determined by the time-independent GPE 
\begin{equation}\label{eq:time_independent_GPE}
 \mu_j' \,\psi^{\text(c)}_j(\x',0) = \left[ \frac{m_j}{2} \, \x'^\tp \Omega^2(0) \, \x' 
 + \sum\limits_{k} g_{j k} \, \big|\psi^{\text(c)}_k(\x',0)\big|^2 \right] \,\psi^{\text(c)}_j(\x',0)
\end{equation}
where $\mu_j'$ is the chemical potential of species $j$
\footnote{The chemical potential $\mu_j'$ is defined here with respect to the potential $\mathcal{V}_j$. Strictly speaking, in general one may need to add $\vec{x}'$-independent terms analogous to those in Eq.~\eqref{eq:phase}, which can be important when the condensate mixture interacts with an external particle reservoir, but are not relevant for our considerations.}.

\subsection{Scaling approach and time-dependent Thomas-Fermi approximation}

In the case of identical central trajectories and locally harmonic potentials described above, the (quantum) dynamics of the centered wave packets, which include the size evolution of the BEC mixture, can be determined quite efficiently without having to solve the non-linear coupled GPE~\eqref{eq:gpe} explicitly.
In fact, for potentials given by Eq.~\eqref{eq:harmonic_potential} the relative shape of the initial density distribution of the mixture is conserved during the time evolution so that for constant interaction strengths the time-dependent wave functions are governed by a single scaling law analogous to the scaling solution for single-species BECs~\cite{Castin1996,Kagan1996,Meister_2017}.

To prove this statement, we introduce the rescaled coordinates
\begin{equation}
  \xxi = \Lambda^{-1}(t) \,\x' ,
\end{equation}
with the time-dependent $3\times 3$ scaling matrix $\Lambda(t)$,
and perform a transformation of the centered wave functions
\begin{equation}\label{eq:rescaled_wavepackets}
  \psi^{\text(c)}_j(\x',t) = \frac{\Me^{\Mi\,\Phi_j(\xxi,t)} }{\sqrt{\det\Lambda(t)}} \, \psi_{\Lambda,j}\left(\xxi, t \right) ,
\end{equation}
with
\begin{equation}\label{eq:transformation:quadratic:phase}
 \Phi_j(\xxi,t) = - \frac{\beta_j (t)}{\hbar} + \frac{m_j}{2 \hbar} (\Lambda\,\xxi )^\tp
 \frac{\Mdiff\Lambda}{\Mdiff t} \, \xxi ,
\end{equation}
and
\begin{equation}
 \beta_j(t) = \int\limits_{t_0}^t \,\Mdiff t' \frac{\mu_j'}{\det \Lambda(t')} .
\end{equation}

By applying this transformation to the coupled GPE~\eqref{eq:gpe} and arranging the terms appropriately, we obtain the rescaled differential equation
\begin{equation}\label{eq:transformed:gpe}
  \Mi \hbar \,\frac{\partial}{\partial t} \,\psi_{\Lambda, j}(\xxi, t)
  = \left[ H_j^{(1)}(\xxi, t) + H_j^{(2)}(\xxi, t)\right] \psi_{\Lambda, j}(\xxi, t) , 
\end{equation}
for the centered wave packets and with the position- and time-depdent terms
\begin{equation}\label{eq:transformed:gpe:H1}
 H_j^{(1)}(\xxi, t) = -\frac{\hbar^2}{2m_j} \nabla_{\xxi}^{\tp} \,\Lambda^{-1} \,\left(\Lambda^{-1}\right)^\tp \,\nabla_{\xxi} ,
 \end{equation}
 and
\begin{equation}\label{eq:transformed:gpe:H2}
 H_j^{(2)}(\xxi, t) = \frac{1}{\det\Lambda} \Big[ \frac{m_j}{2} \, \xxi^\tp \Omega^2(0) \, \xxi
 + \sum\limits_{k} g_{j k} \, \big|\psi_{\Lambda, k}(\xxi, t)\big|^2  - \mu_j' \Big].
\end{equation}
Up to this point the rescaled GPE~\eqref{eq:transformed:gpe} is exact as long as the scaling matrix $\Lambda = \Lambda(t)$ fulfills the ordinary differential equation
\begin{equation}
 \frac{\Mdiff^2\Lambda}{\Mdiff t^2} + \Omega^2(t) \, \Lambda = \frac{\left(\Lambda^{-1}\right)^\tp \Omega^2(0)}{\det\Lambda}
\end{equation}
with the initial conditions
\begin{equation}\label{eq:initial:conditions:Lambda}
 \Lambda(0) = \mathbb{1} \qquad \mathrm{and} \qquad \left.\frac{\Mdiff \Lambda}{\Mdiff t}\right|_{t=0} = 0 \,.
\end{equation}
Clearly, the above transformation requires identical central trajectories and a common frequency matrix $\Omega^2(t)$ for all species, so that a single scaling matrix $\Lambda(t)$ is sufficient to capture the evolution of the entire mixture. 
In the usual case of different trap frequencies for the atomic species, different scalings are necessary for the individual species, which leads to approximate solutions for the mixture dynamics as discussed in Ref.~\cite{Corgier_2020}. In particular, the miscible and immiscible regions of the mixture would expand differently for unequal trap frequencies and give rise to shape deformations, whereas in the case of a common frequency matrix the whole mixture evolves uniformly.

Indeed, as we will show in the following, there is nearly no dynamics of the centered wave packets in the rescaled coordinates and the time evolution is almost entirely determined by the transformation~\eqref{eq:rescaled_wavepackets}. 
First, the contribution of the kinetic term $H_j^{(1)}$, defined in Eq.~\eqref{eq:transformed:gpe:H1}, can be neglected based on the well-known time-dependent Thomas-Fermi approximation~\cite{Dalfovo1999} which is valid as long as the density distributions only undergo spatial changes on length scales larger then the healing length. For the parameters of typical experiments discussed here this assumption is already fulfilled for the ground state of the mixture and stays valid during the time evolution since most of the dynamics is included in the quadratic phase~\eqref{eq:transformation:quadratic:phase}.

Next, one can show that the remaining differential equation, governed by the term $H_j^{(2)}$ defined in Eq.~\eqref{eq:transformed:gpe:H2}, conserves the density $|\psi_{\Lambda,j}(\xxi,t)|^2$ over time because $H_j^{(2)}$ is real and acts only on position space. Thus, we obtain 
\begin{equation}
 |\psi_{\Lambda,j}(\xxi,t)|^2 \approx |\psi_{\Lambda,j}(\xxi,0)|^2 
\end{equation}
and the rescaled GPE~\eqref{eq:transformed:gpe} can be written as
\begin{equation}\label{eq:reduced_transformed_gpe}
 i \hbar \,\frac{\partial}{\partial t} \,\psi_{\Lambda, j}(\xxi, t)
  \approx \frac{1}{\det\Lambda} \Big[ \frac{m_j}{2} \, \xxi^\tp \Omega^2(0) \, \xxi
 + \sum\limits_{k} g_{j k} \, \big|\psi_{\Lambda, k}(\xxi, 0)\big|^2  - \mu_j' \Big]\,\psi_{\Lambda, j}(\xxi, t).
\end{equation}

Since due to the initial conditions~\eqref{eq:initial:conditions:Lambda} we obtain the relation $|\psi^{\text(c)}_j(\x',0)|^2 = |\psi_{\Lambda,j}\left(\xxi, 0 \right)|^2$, we can make use of the time-independent GPE~\eqref{eq:time_independent_GPE} to show that the right hand side of the differential equation~\eqref{eq:reduced_transformed_gpe} vanishes:
\begin{equation}
    \Mi \hbar \,\frac{\partial}{\partial t} \,\psi_{\Lambda,j}(\xxi, t) \approx 0
\end{equation}

Hence, the dynamics in the adapted coordinates is practically frozen: $\psi_{\Lambda,j}(\xxi,t) \approx \psi_{\Lambda,j}(\xxi,0) \approx \psi^{\text(c)}_j(\x',0)$. Therefore, the dynamics in the comoving frame is given by the initial density profile $\psi^{\text(c)}_j(\x',0)$ and the transformation~\eqref{eq:rescaled_wavepackets} determined by the scaling matrix $\Lambda(t)$.

\end{document}